\documentstyle[12pt]{article}
\newlength{\extraspace}
\newlength{\extraspaces}
\newcommand{\be}{\begin{equation}
\addtolength{\abovedisplayskip}{\extraspaces}
\addtolength{\belowdisplayskip}{\extraspaces}
\addtolength{\abovedisplayshortskip}{\extraspace}
\addtolength{\belowdisplayshortskip}{\extraspace}}
\newcommand{\ee}{\end{equation}}

\newcommand{\ba}{\begin{eqnarray}
\addtolength{\abovedisplayskip}{\extraspaces}
\addtolength{\belowdisplayskip}{\extraspaces}
\addtolength{\abovedisplayshortskip}{\extraspace}
\addtolength{\belowdisplayshortskip}{\extraspace}}
\newcommand{\ea}{\end{eqnarray}}

\newcommand{\newsection}[1]{
\vspace{15mm}
\pagebreak[3]
\addtocounter{section}{1}
\setcounter{equation}{0}
\setcounter{subsection}{0}
\setcounter{footnote}{0}
\begin{flushleft}
{\large\bf \thesection. #1}
\end{flushleft}
\nopagebreak
\medskip
\nopagebreak}

\newcommand{\Tr}{{\rm Tr}}
\newcommand{\Dmrns}{{\cal D}_{\mu\rho,\nu\sigma}}

\begin{document}

\addtolength{\baselineskip}{.8mm}

{\thispagestyle{empty}
\noindent \hspace{1cm}  \hfill March 1996 \hspace{1cm}\\
\mbox{}                 \hfill IFUP--TH 12/96 \hspace{1cm}\\
\mbox{}                 \hfill UCY--PHY--96/5 \hspace{1cm}\\

\begin{center}\vspace*{1.0cm}
{\large\bf Field strength correlations in the}\\
{\large\bf QCD vacuum at short distance
 \footnote{Partially supported by MURST (Italian Ministry of the University 
 and of Scientific and Technological Research) and by the EC contract 
 CHEX--CT92--0051.} }\\
\vspace*{1.0cm}
{\large A. Di Giacomo, E. Meggiolaro}\\
\vspace*{0.5cm}{\normalsize
{Dipartimento di Fisica, \\
Universit\`a di Pisa, \\ 
and INFN, Sezione di Pisa,\\
I--56100 Pisa, Italy.}}\\
\vspace*{1.0cm}
{\large H. Panagopoulos}\\
\vspace*{0.5cm}{\normalsize
{Department of Natural Sciences,\\
University of Cyprus,\\
1678 Nicosia, Cyprus.}}\\
\vspace*{2cm}{\large \bf Abstract}
\end{center}
\noindent
We determine by numerical simulations on a lattice the gauge--invariant
two--point correlation function of the gauge field strengths in the QCD 
vacuum, down to a distance of 0.1 fm.
}
\vfill\eject

\newsection{Introduction}

\noindent
The gauge--invariant two--point correlators of the field strengths in the 
QCD vacuum are defined as
\be
\Dmrns(x) = \langle 0| 
\Tr \left\{ G_{\mu\rho}(x) S(x,0) G_{\nu\sigma}(0) S^\dagger(x,0) \right\}
|0\rangle ~,
\ee
where $G_{\mu\rho} = gT^aG^a_{\mu\rho}$ and $T^a$ are the generators of the 
colour gauge group in the fundamental representation. 
Moreover, in Eq. (1.1),
\be
S(x,0) = {\rm P}\exp\left(i\int^1_0dt\,x^\mu A_\mu(xt)\right) ~,
\ee
with $A_\mu=gT^aA^a_\mu$, is the Schwinger phase operator needed to 
parallel--transport the tensor $G_{\nu\sigma}(0)$ to the point $x$.

These field--strength correlators play an important role in hadron physics.
In the spectrum of heavy $Q \bar{Q}$ bound states, they govern the effect 
of the gluon condensate on the level splittings 
\cite{Gromes82,Campostrini86,Simonov95}.
They are the basic quantities in models of stochastic confinement of colour
\cite{Dosch87,Dosch88,Simonov89}
and in the description of high--energy hadron scattering
\cite{Nachtmann84,Landshoff87,Kramer90,Dosch94}.

A numerical determination of the correlators on lattice (with gauge group
$SU(3)$) already exists, in the range of physical distances between 0.4 and 1 
fm \cite{DiGiacomo92}. In that range $\Dmrns$ falls off exponentially 
\be
\Dmrns(x) \sim \exp(-|x|/\lambda) ~,
\ee
with a correlation length $\lambda \simeq 0.22$ fm \cite{DiGiacomo92}.

What makes the determination of the correlators possible on the lattice,
with a reasonable computing power, is the idea \cite{Campostrini89,DiGiacomo90}
of removing the effects of short--range fluctuations on 
large distance correlators by a local {\it cooling} procedure.
Freezing the links of QCD configurations one after the other, damps very 
rapidly the modes of short wavelength, but requires a number $n$ of cooling 
steps proportional to the square of the distance $d$ in lattice units to 
affect modes of wavelength $d$:
\be
n \simeq k d^2 ~.
\ee
Cooling is a kind of diffusion process.
If $d$ is sufficiently large, there will be a range of values of $n$ in 
which lattice artefacts due to short--range fluctuations have been removed, 
without touching the physics at distance $d$; by {\it lattice artefacts} we 
mean statistical fluctuations and renormalization effects from lattice to 
continuum. This removal will show up as a plateau in the dependence 
of the correlators on $n$. This was the technique successfully used in
Ref. \cite{DiGiacomo92}. There, the range of distances explored was from 
from 3--4 up to 7--8 lattice spacings at $\beta \simeq 6.$, which means
approximately from 0.4 up to 1 fm in physical distance.
The lattice size was $16^4$.

We have now new results on a $32^4$ lattice, at $\beta$
between 6.6 and 7.2: at these values of $\beta$ the lattice size is still 
bigger than 1 fm, and therefore safe from infrared artefacts, but
$d = 3,4$ lattice spacings now correspond to physical distances of about
0.1 fm. Since what matters to our cooling procedure is the distance in 
lattice units, we obtain in this way a determination of the correlators at 
distances down to 0.1 fm.

\newsection{Computations and results}

\noindent
The most general form of the correlator compatible with the invariances of 
the system is \cite{Dosch87,Dosch88,Simonov89}
\ba
\lefteqn{
\Dmrns(x) = (g_{\mu\nu}g_{\rho\sigma} - g_{\mu\sigma}g_{\rho\nu})
\left[ {\cal D}(x^2) + {\cal D}_1(x^2) \right] } \nonumber \\
& & + (x_\mu x_\nu g_{\rho\sigma} - x_\mu x_\sigma g_{\rho\nu} 
+ x_\rho x_\sigma g_{\mu\nu} - x_\rho x_\nu g_{\mu\sigma})
{\partial{\cal D}_1(x^2) \over \partial x^2} ~.
\ea
${\cal D}$ and ${\cal D}_1$ are invariant functions of $x^2$. We work in 
the Euclidean region.

We can define a ${\cal D}_\parallel(x^2)$ and a ${\cal D}_\perp(x^2)$ as 
follows. We go to a reference frame in which $x^\mu$ is parallel to one of 
the coordinate axes, say $\mu=0$. Then
\ba
{\cal D}_\parallel &\equiv& {1\over3} \sum_{i=1}^3
{\cal D}_{0i,0i}(x) = {\cal D} + {\cal D}_1 + x^2 {\partial{\cal D}_1
\over \partial x^2} ~, \nonumber \\
{\cal D}_\perp &\equiv& {1\over3}\sum_{i<j=1}^3
{\cal D}_{ij,ij}(x) = {\cal D} + {\cal D}_1 ~.
\ea
On the lattice we can define a lattice operator $\Dmrns^L$, which is
proportional to $\Dmrns$ in the continuum limit, i.e., when the lattice 
spacing $a \to 0$. Since the lattice analogue of the field strength is the 
plaquette $\Pi_{\mu\rho}(n)$ (the parallel transport along an elementary 
square of the lattice, lying on the $\mu\rho$--plane), $\Dmrns^L$ will be 
defined as
\be
\Dmrns^L(\hat d a) = {1\over2} \Re \left\{ \langle \Tr [
\Pi_{\mu\rho}(n+\hat d a) S(n+\hat d a,n) 
\Omega_{\nu\sigma}(n) S^\dagger(n+\hat d a,n) ] \rangle 
\right\} ~,
\ee
where $\Re$ stands for real part and 
the lattice operator $\Omega_{\nu\sigma}(n)$ is given by
\be
\Omega_{\nu\sigma}(n) =
\Pi^\dagger_{\nu\sigma}(n) - \Pi_{\nu\sigma}(n)
-{1 \over 3} \Tr [\Pi^\dagger_{\nu\sigma}(n) - \Pi_{\nu\sigma}(n)] ~.
\ee
$\hat d a$ is a line parallel to one of the coordinate axes, 
with integer length $d$ in units of the lattice spacing $a$, joining the two 
sites to be correlated; $S$ is the parallel transport along this line. 
The inclusion of the operator ${1 \over 2} \Omega_{\nu\sigma}$ (in place of
simply putting $\Pi^\dagger_{\nu\sigma}$) on the right--hand side of Eq. (2.3) 
ensures that the {\it disconnected} part and the {\it singlet} part of the 
correlator are left out. In particular, the subtraction of the {\it 
singlet} part (which is anyway a small contribution of order 
${\cal O}(a^{12})$) means that we are indeed taking the correlation of two 
operators with the quantum numbers of a colour octet.

In the na\"\i ve continuum limit ($a \to 0$) we have that
\be
\Dmrns^L(\hat d a) \mathop\sim_{a\to0} a^4 \Dmrns(\hat d a) + 
{\cal O}(a^6) ~. 
\ee
Making use of the definition (2.1) we can also write, in the same limit,
\ba
{\cal D}_\parallel^L(\hat d a) \mathop\sim_{a\to0} a^4 
{\cal D}_\parallel(d^2 a^2) + {\cal O}(a^6) ~,\nonumber \\
{\cal D}_\perp^L(\hat d a) \mathop\sim_{a\to0} a^4 
{\cal D}_\perp(d^2 a^2) + {\cal O}(a^6) ~.
\ea
Equations (2.5) and (2.6) 
come from a formal expansion of the operator, and are expected to be modified, 
when the expectation value is computed, by lattice artefacts, i.e., by 
effects due to the ultraviolet {\it cutoff}. These effects can be estimated in 
perturbation theory and subtracted \cite{Campostrini84}. 
Instead we remove them by freezing the quantum fluctuations at the scale of 
the lattice spacing. After cooling, ${\cal D}_\parallel^L$ and 
${\cal D}_\perp^L$ are expected to obey Eq. (2.6).

We shall omit the details of our cooling procedure, which have been 
described in several papers (see Refs. \cite{Campostrini89,DiGiacomo90} 
and references therein). 
We only remark that it is a local procedure, which affects
correlations at distances that grow with the number of cooling steps as in 
a diffusion process. We then expect that, if the distance at which we 
observe the correlation is sufficiently large, lattice artefacts are frozen 
by cooling long before the correlation is affected: this produces a 
plateau in the correlation versus cooling step. 
Our data are the values of the correlation at the plateau; 
the error is the typical statistical error at the plateau, 
{\it plus} a systematic error which is estimated as the difference 
between neighbouring points at the plateau (whenever the 
plateau is not long enough, looking more like an extremum). 
Typically the global error is three times larger than the statistical 
error. 
The typical behaviour of ${\cal D}_\parallel^L$ and of ${\cal D}_\perp^L$ 
along cooling is shown in Fig. 1.

We have measured the correlations on a $32^4$ lattice at distances ranging 
from 3 to 14 lattice spacings and at $\beta = 6.6,~6.8,~7.0,~7.2$. From 
renormalization group arguments,
\be
a = {1\over\Lambda_L} f(\beta) ~.
\ee
As $\beta \to \infty$, $f(\beta)$ is given by
\be
f(\beta) = \left({8\over33}\,\pi^2\beta\right)
^{ 51/121 } \exp\left(-{4\over33}\pi^2\beta\right)
\left[1+{\cal O}(1/\beta)\right] ~,
\ee
for gauge group $SU(3)$ and in the absence of quarks.
At sufficiently large $\beta$ one expects that
\ba
{\cal D}_\parallel^L f(\beta)^{-4} &=& {1\over\Lambda_L^4}
{\cal D}_\parallel\left({d^2\over\Lambda_L^2}f^2(\beta)\right) ~,
\nonumber \\
{\cal D}_\perp^L f(\beta)^{-4} &=& {1\over\Lambda_L^4}
{\cal D}_\perp\left({d^2\over\Lambda_L^2}f^2(\beta)\right) ~,
\ea
where $f(\beta)$ is given by Eq. (2.8) and terms of higher order in $a$ are 
negligible.

In Figs. 2 and 3 we plot respectively ${\cal D}_\parallel^L f(\beta)^{-4}$ 
and ${\cal D}_\perp^L f(\beta)^{-4}$ versus $d_{\rm phys} = 
(d/\Lambda_L)$  $f(\beta)$. In these figures we have also plotted the values 
of the correlators obtained in Ref. \cite{DiGiacomo92}, corresponding to 
physical distances $d_{\rm phys} \ge 0.4$ fm.
We have applied a best fit to all of these data with the functions
\ba
{\cal D}(x^2) &=& A \exp\left(-{|x|\over\lambda_A}\right)
+ {a\over|x|^4} \exp\left(-{|x|\over\lambda_a}\right) ~,
\nonumber \\
{\cal D}_1(x^2) &=& B \exp\left(-{|x|\over\lambda_A}\right)
+{b\over|x|^4} \exp\left(-{|x|\over\lambda_a}\right) ~.
\ea
We have obtained the following results:
\ba
{A \over \Lambda_L^4} \simeq 3.3 \times 10^8 &~,~&
{B \over \Lambda_L^4} \simeq 0.7 \times 10^8 ~, \nonumber \\
a \simeq 0.69 &~,~& b \simeq 0.46 ~, \nonumber \\
\lambda_A \simeq {1\over\Lambda_L}{1\over182} &~,~&
\lambda_a \simeq {1\over\Lambda_L}{1\over94} ~,
\ea
with $\chi^2/N_{\rm d.o.f.} \simeq 1.7$. The continuum lines in Figs. 2 and 
3 have been obtained using the parameters of this best fit.
With the value of $\Lambda_L$
determined from the string tension \cite{Michael88} we obtain
\be
\lambda_A \simeq 0.22 \, {\rm fm} ~,~ \lambda_a \simeq 0.43 \, {\rm fm} ~.
\ee
The correlation length $\lambda_A$, which enters the non--perturbative
exponential terms of ${\cal D}$ and ${\cal D}_1$, as well as the magnitude 
of the coefficients $A$ and $B$, are compatible with the values obtained
in Ref. \cite{DiGiacomo92}.
We have also tried a different fit, in which ${\cal D}_1$ is described by a 
purely perturbative--like term, while ${\cal D}$ is still the sum of a 
non--perturbative exponential term {\it plus} a purely perturbative--like 
term. In other words, we have fixed $B = 0$ and $\lambda_a = 0$ in the 
parametrization (2.10) for ${\cal D}$ and ${\cal D}_1$:
\ba
{\cal D}(x^2) &=& A \exp\left(-{|x|\over\lambda_A}\right)
+ {a\over|x|^4} ~, \nonumber \\
{\cal D}_1(x^2) &=& {b\over|x|^4} ~.
\ea
We have thus obtained:
\ba
{A \over \Lambda_L^4} \simeq 2.7 \times 10^8 &~,~&
\lambda_A \simeq {1\over\Lambda_L}{1\over183} ~, \nonumber \\
a \simeq 0.4 &~,~& b \simeq 0.3 ~.
\ea
The value of $\chi^2/N_{\rm d.o.f.}$ is again acceptable (about 2); the slope 
of the non--perturbative exponential term is practically unchanged.
The behaviour for ${\cal D}_\parallel$ and ${\cal D}_\perp$ obtained using 
the parameters from this best fit is close to the continuum lines reported 
in Figs. 2 and 3.

As a final comment we notice that we have been able to observe terms 
proportional to $1/|x|^4$ in the correlations because we have worked at 
larger values of $\beta$, where the distance between two points 
(far enough in lattice units so that the correlation is not modified by 
cooling before lattice artefacts are eliminated) is small compared with 
$1\,{\rm fm}$ in physical units. A larger lattice ($32^4$) has been 
necessary to avoid infrared artefacts.

\bigskip
\noindent {\bf Acknowledgements}
\smallskip

This work was done using the CRAY T3D of the CINECA Inter University 
Computing Centre (Bologna, Italy). We would like to thank the CINECA for 
having put the CRAY T3D at our disposal and for the kind and highly qualified 
technical assistance. 

We thank G\"unther Dosch and Yuri Simonov for many 
useful discussions.

\vfill\eject

{\renewcommand{\Large}{\normalsize}
}

\vfill\eject

\noindent
\begin{center}
{\bf FIGURE CAPTIONS}
\end{center}
\vskip 0.5 cm
\begin{itemize}
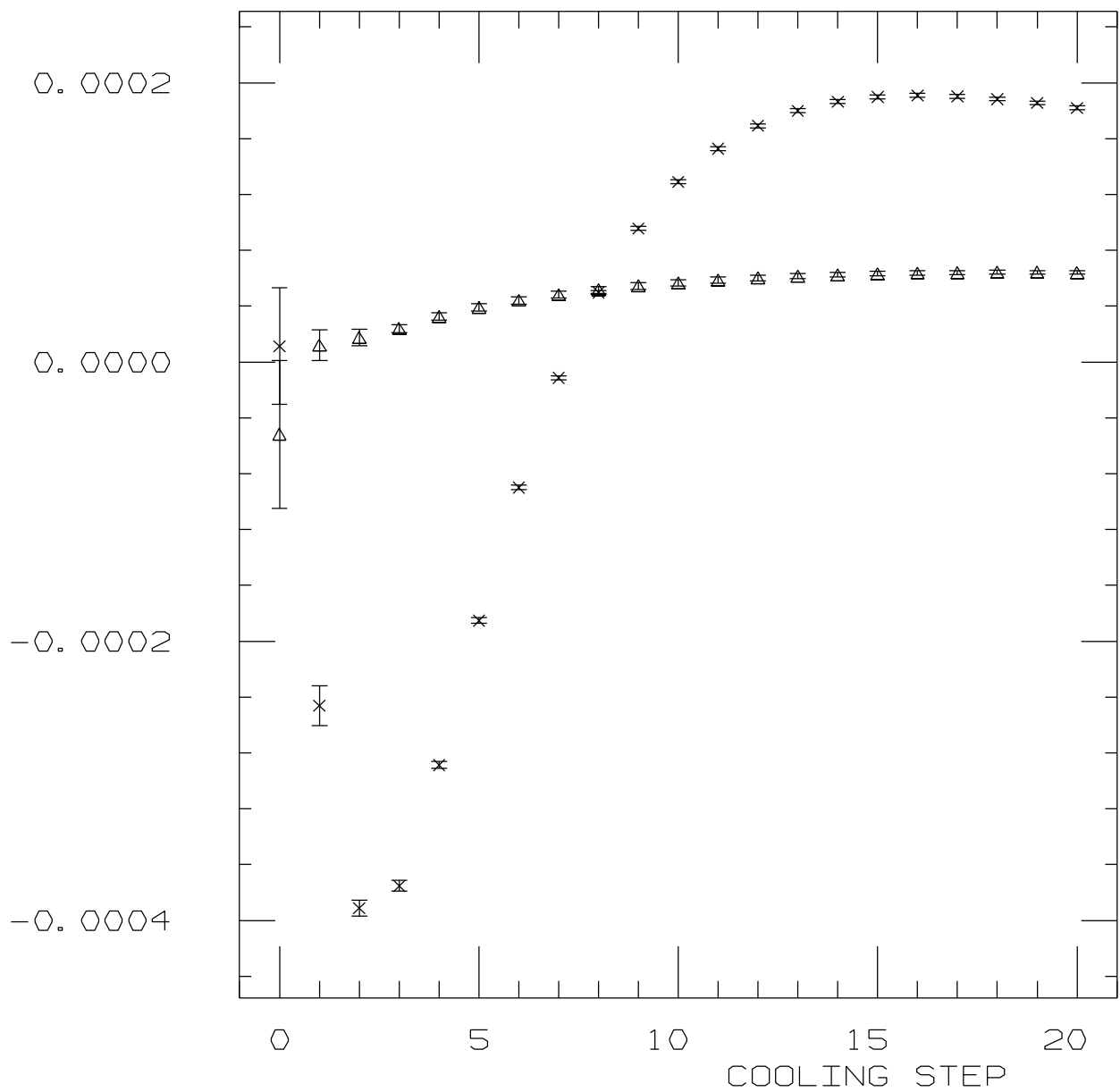
\item [\bf Fig.~1.] A typical behaviour of ${\cal D}_\parallel^L$
(crosses; $d = 6$, $\beta = 6.6$, lattice $32^4$)
and of ${\cal D}_\perp^L$ (triangles; $d = 12$, $\beta = 6.6$, lattice $32^4$)
during cooling. 
\bigskip
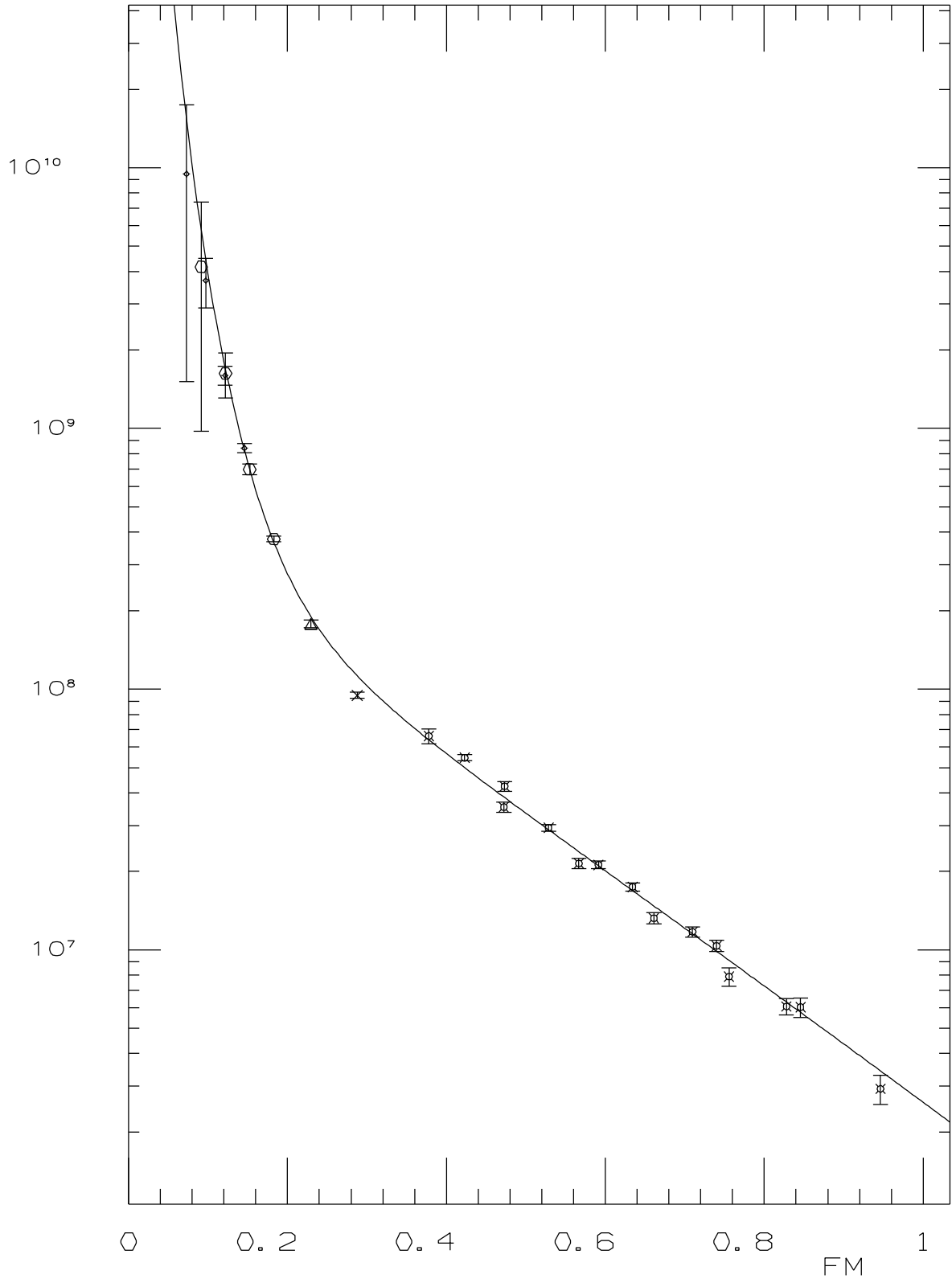
\item [\bf Fig.~2.] The function ${\cal D}_\parallel^L f(\beta)^{-4}$ 
versus physical distance (in {\it fermi} units). Crosses
correspond to $\beta=6.6$, triangles to $\beta=6.8$, hexagons to $\beta=7.0$, 
diamonds to $\beta=7.2$; crossed--circles correspond to the data of 
Ref. \cite{DiGiacomo92}. The line is the curve for ${\cal D}_\parallel$
obtained from the best fit of Eqs. (2.10) and (2.11).
\bigskip
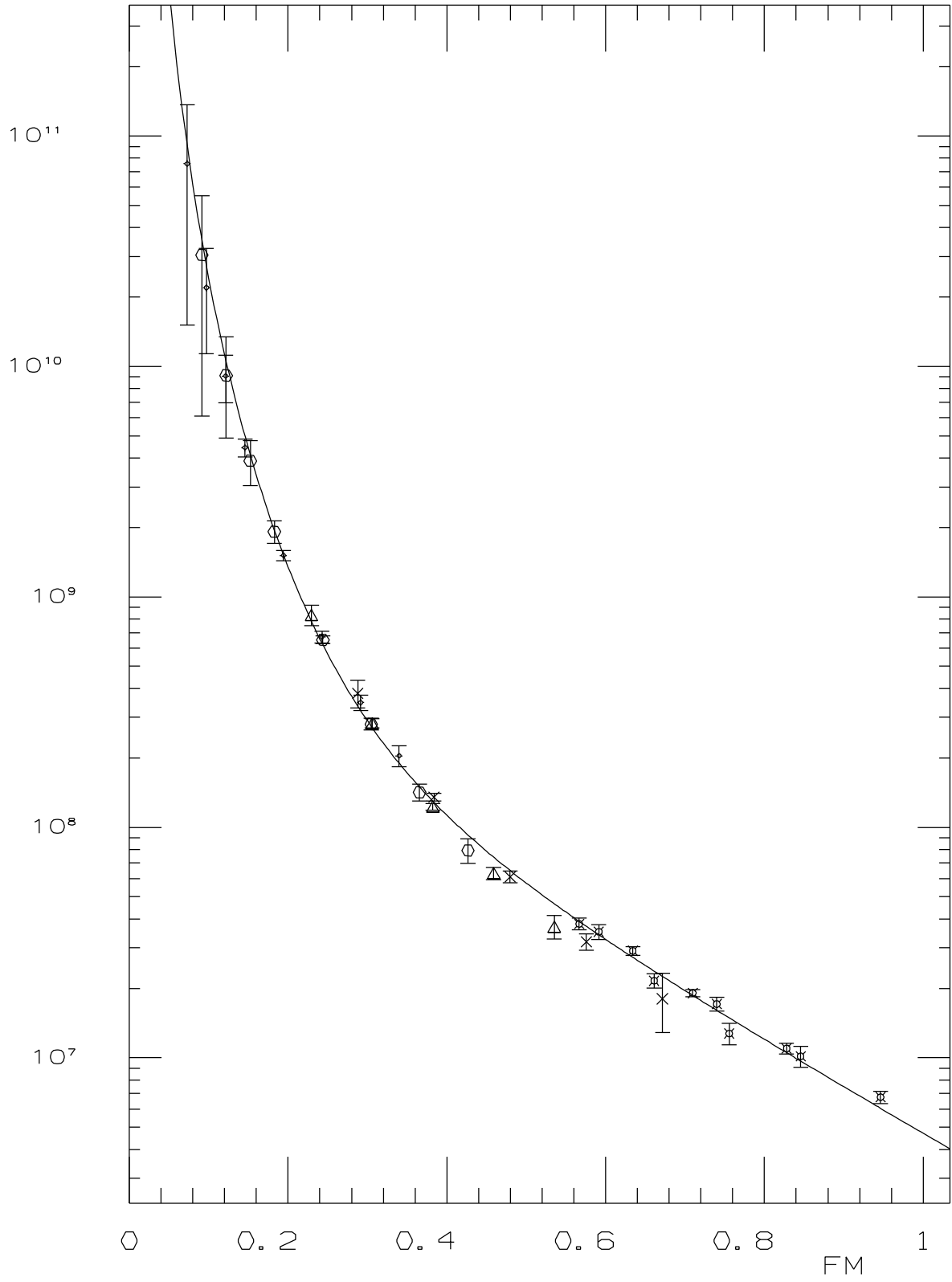
\item [\bf Fig.~3.] The function ${\cal D}_\perp^L f(\beta)^{-4}$ 
versus physical distance (in {\it fermi} units). The symbols are the same 
as in Fig. 2. The line is the curve for ${\cal D}_\perp$
obtained from the best fit of Eqs. (2.10) and (2.11).
\end{itemize}

\vfill\eject

\end{document}